\documentclass[11pt]{article}
\usepackage[utf8]{inputenc}
\usepackage[T1]{fontenc}
\usepackage{amsmath, amssymb, amsfonts, bm}
\usepackage{slashed}
\usepackage{graphicx}
\usepackage{geometry}
\usepackage{cite}
\usepackage{hyperref}
\usepackage{booktabs}

\usepackage{subcaption}
\usepackage[compat=1.1.0]{tikz-feynman}

\usepackage{authblk}

\title{\textbf{Probing the Tau Anomalous Magnetic Moment at Colliders: From Ultra-Peripheral Collisions to the Precision Frontier}}
\author{Natascia Vignaroli}
\affil{Dipartimento di Matematica e Fisica, Università del Salento, and Istituto Nazionale di Fisica
Nucleare, Sezione di Lecce, I-73100 Lecce, Italy}

\date{\today}

\begin{document}

\maketitle


\thispagestyle{empty}

\begin{abstract}
The anomalous magnetic moment of the tau lepton, $a_{\tau}$, represents a fundamental test of the Standard Model (SM) and a high-sensitivity probe for New Physics in the third generation of leptons. Due to the tau's extremely short lifetime, traditional spin-precession measurements remain inaccessible, necessitating innovative experimental strategies at high-energy colliders. This review provides a comprehensive overview of the current experimental landscape, highlighting the recent paradigm shift from LEP-era constraints to the unprecedented precision reached at the LHC. We emphasize the importance of Ultra-Peripheral Heavy-Ion Collisions (UPCs), which act as a ``photon-photon collider'' of extreme intensity. By leveraging the $Z^4$ enhancement of the coherent photon flux in Lead-Lead ($PbPb$) interactions, these collisions provide a theoretically robust ``quasi-static'' environment. These results are critically compared with the latest measurements from proton-proton collisions, including the recent CMS observation of the $\gamma\gamma \to \tau\tau$ process and the ATLAS constraints from the high-mass Drell-Yan tail. We evaluate their complementarity and the challenges related to Effective Field Theory validity at the TeV scale. Finally, we outline the future prospects for $a_\tau$ at Belle II and the Future Circular Collider (FCC) stages. While FCC-hh in $PbPb$ mode provides a theoretically clean environment, its sensitivity remains limited to $\mathcal{O}(10^{-2})$. Conversely, the next generation of lepton facilities, specifically Belle II and FCC-ee, aims for the $\mathcal{O}(10^{-5})$ level, required to probe SM electroweak loop corrections. Long-term projections for a high-energy Muon Collider suggest a potential reach of $\mathcal{O}(10^{-6})$.
\end{abstract}

\newpage

\tableofcontents

\section{Introduction and Theoretical Framework} 

The anomalous magnetic moment of the tau lepton, $a_\tau = (g_\tau - 2)/2$, represents an important probe of the Standard Model and its extensions. While the electron and muon anomalies have been measured with staggering precision, the tau remains comparatively less constrained experimentally due to its short lifetime ($\approx 290$ fs), which prevents the use of traditional storage ring techniques. However, the tau's large mass ($m_\tau \approx 1777$ MeV) provides a unique advantage: it can enhance the sensitivity to certain classes of New Physics. Since many Beyond the Standard Model (BSM) effects scale with the square of the lepton mass, in scenarios where the chirality flip is proportional to the lepton mass $a_\tau$ is potentially more sensitive by a factor of $m^2_\tau/m^2_\mu \approx 280$ in scenarios where dipole operators scale with the lepton mass, making it a primary candidate for testing Lepton Flavor Universality and New Physics in general. \\

The electromagnetic properties of the tau lepton are fundamentally encoded in the vertex function $\Gamma^\mu(q)$, which describes the interaction between the leptonic current and the photon field. In the simplest case of the Dirac theory, the tau is treated as a point-like particle, leading to the tree-level prediction of the gyromagnetic ratio $g^{Classic}_\tau = 2$. However, quantum field theory implies that this ``bare'' vertex is modified by radiative corrections, necessitating a more general parametrization.

The structure of $\Gamma^\mu$ is constrained by the $U(1)_Q$ gauge symmetry of the Standard Model. For an on-shell tau lepton with incoming momentum $p$ and outgoing momentum $p'$, the momentum transfer is defined as $q = p' - p$. Gauge invariance requires the electromagnetic current to be conserved, leading to the Ward identity
$q_\mu \Gamma^\mu(q) = 0$ for on-shell external fermions. This identity restricts the allowed Lorentz structures.
By applying the Gordon decomposition and respecting Lorentz covariance, the most general vertex function for a spin-1/2 on-shell lepton can be expressed through four independent, dimensionless form factors $F_i(q^2)$:

\begin{equation}\label{eq:vertex}
\Gamma^\mu(q) = -ie \left[ F_1(q^2) \gamma^\mu + \frac{i\sigma^{\mu\nu}q_{\nu}}{2m_\tau} F_2(q^2) + \frac{\sigma^{\mu\nu}q_{\nu}\gamma_5}{2m_\tau} F_3(q^2) + \frac{F_A(q^2)}{m^2_\tau}(q^2\gamma^\mu - q^\mu \slashed{q})\gamma_5  \right]
\end{equation}

where $\sigma^{\mu\nu}=\frac{i}{2}[\gamma^\mu, \gamma^\nu]$. 
The last term in Eq.~(1) defines the anapole form factor $F_A(q^2)$, which is associated with a parity-violating but time-reversal conserving interaction. Its Lorentz structure, proportional to $(q^2 \gamma^\mu - q^\mu \slashed{q})\gamma^5$, is uniquely fixed by electromagnetic gauge invariance and ensures current conservation, $q_\mu \Gamma^\mu = 0$. Unlike the Dirac and Pauli form factors, the anapole contribution vanishes for on-shell photons ($q^2 =0$), and therefore does not contribute to static electromagnetic properties such as the electric charge or the anomalous magnetic moment. Instead, it parameterizes momentum-dependent effects that can only be probed in processes involving off-shell photons. For this reason, it will not be relevant for the extraction of $a_\tau$ discussed in this work.
In some conventions, an explicit factor $1/q^2$ is introduced in the definition of the anapole term so that $F_A(0)$ directly represents a static anapole moment. In the present work, we adopt the convention of Eq.~\eqref{eq:vertex}, where the form factor remains regular and the anapole contribution vanishes in the static limit.

The physical interpretation of the form factors is obtained in the static limit ($q^2 \to 0$):

\begin{itemize}
    \item \textbf{$F_1(0) = 1$}: This is the Dirac form factor. Its value is fixed by the renormalization of the electric charge. It represents the minimal coupling where the lepton interacts as a point-like charge $e$.
    \item \textbf{$F_2(0) = a_\tau$}: This is the Pauli form factor, defining the Anomalous Magnetic Moment. Since $F_2$ is identically zero at tree-level in the Dirac equation, any non-zero value arises exclusively from loop-level radiative corrections.
    \item \textbf{$F_3(0) = \frac{2m_\tau}{e} d_\tau$} in the commonly adopted convention: This factor defines the Electric Dipole Moment (EDM). Unlike the previous terms, $F_3$ is associated with an operator that violates both Parity (P) and Time-Reversal (T) symmetries. Under the assumption of CPT conservation, this term is a direct probe of CP violation in the leptonic sector.
\end{itemize}

The transition from the general vertex structure to the observable anomalous magnetic moment is established in the static limit. By mapping the effective interaction onto the non-relativistic Hamiltonian for a generic spin-1/2 particle of mass $m$ in a magnetic field, the gyromagnetic ratio $g$, which relates the magnetic moment and the particle spin as $\vec{\mu} = g \left( \frac{e}{2m} \right) \vec{S}$, is found to be:
\begin{equation}
g = 2 [F_1(0) + F_2(0)]
\end{equation}
Given that $F_1(0) = 1$ due to charge renormalization, the Dirac equation's prediction of $g=2$ is modified by the Pauli form factor $F_2(0)$. This defines the anomalous magnetic moment as 
\begin{equation}
a = (g - 2)/2 = F_2(0)
\end{equation}

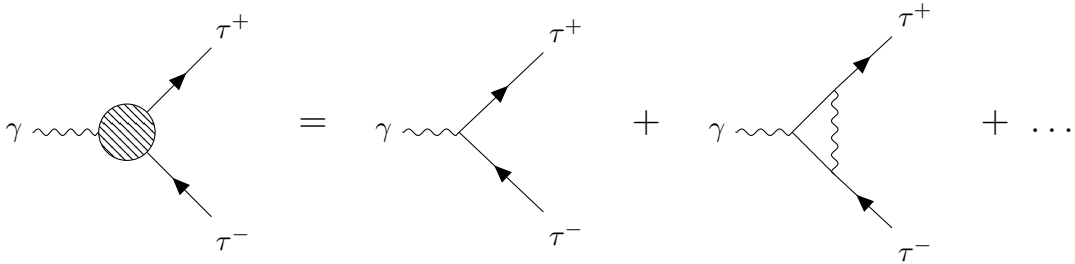
\begin{figure}[h]
    \centering
    \begin{tikzpicture}[baseline=(b.base)]
      \begin{feynman}
        \vertex (a) {\(\gamma\)};
        \vertex [right=1.5cm of a, blob] (b) {};
        \vertex [above right=2.cm of b] (f1) {\(\tau^{+}\)};
        \vertex [below right=2.cm of b] (f2) {\(\tau^{-}\)};
        \diagram* {
          (a) -- [photon] (b),
          (b) -- [fermion] (f1),
          (f2) -- [fermion] (b),
        };
      \end{feynman}
    \end{tikzpicture}
    \quad \textbf{\Large \( = \)} \quad
    \begin{tikzpicture}[baseline=(b.base)]
      \begin{feynman}
        \vertex (a) {\(\gamma\)};
        \vertex [right=1cm of a] (b);
        \vertex [above right=1.5cm of b] (f1) {\(\tau^{+}\)};
        \vertex [below right=1.5cm of b] (f2) {\(\tau^{-}\)};
        \diagram* {
          (a) -- [photon] (b),
          (b) -- [fermion] (f1),
          (f2) -- [fermion] (b),
        };
      \end{feynman}
    \end{tikzpicture}
    \quad \textbf{\Large \( + \)} \quad
    \begin{tikzpicture}[baseline=(b.base)]
      \begin{feynman}
        \vertex (a) {\(\gamma\)};
        \vertex [right=1cm of a] (v_ext);
        \vertex [above right=0.8cm of v_ext] (v1);
        \vertex [below right=0.8cm of v_ext] (v2);
        \vertex [above right=1.cm of v1] (f1) {\(\tau^{+}\)};
        \vertex [below right=1.cm of v2] (f2) {\(\tau^{-}\)};
        \diagram* {
          (a) -- [photon] (v_ext),
          (v_ext) -- [plain] (v1), 
          (v2) -- [plain] (v_ext), 
          (v1) -- [fermion] (f1),  
          (f2) -- [fermion] (v2),  
          (v1) -- [photon] (v2),   
        };
      \end{feynman}
    \end{tikzpicture}
    \quad \textbf{\Large \( + \ \dots \)}
    
    \vspace{0.5cm}
    \caption{Decomposition of the dressed $\tau\tau\gamma$ vertex. The physical vertex (left) is the sum of the Dirac tree-level contribution and the 1-loop Schwinger correction (with internal photon and fermion lines) plus higher-order terms.}\label{fig:effective-vertex}
\end{figure}

Thus, $a_\tau$ encapsulates all radiative corrections that shift the tau lepton from its point-like behavior (Fig. \ref{fig:effective-vertex}).

The anomalous magnetic moment $a_\tau$ thus arises exclusively from quantum fluctuations beyond the tree-level Dirac prediction. The dominant contribution is the universal first-order QED correction, calculated by Schwinger in 1948 as $a_\tau^{(1)} = \alpha/2\pi \approx 1161.4 \times 10^{-6}$. However, a full Standard Model assessment requires the inclusion of higher-order QED terms, electroweak loops, and hadronic effects.

According to the comprehensive assessment by Eidelman and Passera \cite{Eidelman:2007sb}, the current SM prediction is:
\begin{equation}
a_\tau^{\text{SM}} = 117\,721 \, (5) \times 10^{-8}
\end{equation}
The total value is decomposed into the following primary sectors:

\begin{itemize}
    \item \textbf{QED Contributions ($a_\tau^{\text{QED}}$):} This sector accounts for the largest fraction of the total value, estimated at $117\,324 \times 10^{-8}$. It includes all loops involving photons and leptons. While the QED terms for the electron and muon are known to extremely high orders, the $a_\tau$ precision is dominated by the leading-order diagrams, with the uncertainty being negligible compared to the hadronic sector.
    
    \item \textbf{Electroweak Contributions ($a_\tau^{\text{EW}}$):} These corrections involve loops with $W^{\pm}$, $Z$, and Higgs bosons, contributing $47.4 \, (0.5) \times 10^{-8}$. The large mass of the tau lepton enhances these effects; specifically, the EW contribution to $a_\tau$ is roughly 400 times larger than its counterpart for the electron and significantly more prominent than for the muon, making the tau an ideal probe for New Physics (NP) scales coupled to the mass of the leptons.
    
    \item \textbf{Hadronic Contributions ($a_\tau^{\text{Had}}$):} Estimated at $350.1 \, (4.8) \times 10^{-8}$, this sector is the dominant source of theoretical uncertainty. It is subdivided into:
    \begin{itemize}
        \item \textbf{Hadronic Vacuum Polarization (HVP):} $a_\tau^{\text{HVP}} = 348.6 \, (4.7) \times 10^{-8}$.
        \item \textbf{Hadronic Light-by-Light (LbL):} $a_\tau^{\text{LbL}} = 1.5 \, (1.0) \times 10^{-8}$.
    \end{itemize}
    These terms involve non-perturbative QCD effects. The HVP contribution is typically evaluated using experimental data from $e^+e^- \to \text{hadrons}$ cross-sections via dispersion relations.
\end{itemize}

The total theoretical uncertainty of $\pm 5 \times 10^{-8}$ is almost entirely dictated by the hadronic sector. 
This theoretical precision provides a well-defined reference prediction for current experimental searches. Any significant deviation observed in collider data would indicate the presence of physics beyond the Standard Model. 

$a_\tau$ could indeed increase significantly if we consider Beyond the Standard Model (BSM) physics, as the tau lepton acts as a natural amplifier for heavy-scale interactions.

\begin{figure}[h]
    \centering
    \begin{minipage}{0.48\textwidth}
        \centering
        \begin{tikzpicture}[scale=0.8]
          \begin{feynman}
            \vertex (a) {\(\gamma\)}; 
            \vertex [right=1.2cm of a] (v);
            \vertex [above right=1.2cm of v] (f1) {\(\tau\)}; 
            \vertex [below right=1.2cm of v] (f2) {\(\tau\)};
            \vertex [above right=0.6cm of v] (v1); 
            \vertex [below right=0.6cm of v] (v2);
            \diagram* { 
                (a) -- [photon] (v), 
                (v) -- [dashed, edge label=\(\tilde{\tau}\)] (v1) -- [fermion] (f1), 
                (f2) -- [fermion] (v2) -- [dashed, edge label=\(\tilde{\tau}\)] (v), 
                (v1) -- [fermion, edge label=\(\tilde{\chi}^0\)] (v2) 
            };
          \end{feynman}
        \end{tikzpicture}
        \subcaption{SUSY contribution}
    \end{minipage}
    \hfill
    \begin{minipage}{0.48\textwidth}
        \centering
        \begin{tikzpicture}[scale=0.8]
          \begin{feynman}
            \vertex (a) {\(\gamma\)}; 
            \vertex [right=1.2cm of a] (v);
            \vertex [above right=1.2cm of v] (f1) {\(\tau\)}; 
            \vertex [below right=1.2cm of v] (f2) {\(\tau\)};
            \vertex [above right=0.6cm of v] (v1); 
            \vertex [below right=0.6cm of v] (v2);
            \diagram* { 
                (a) -- [photon] (v), 
                (v) -- [fermion, edge label=\(q\)] (v1) -- [fermion] (f1), 
                (f2) -- [fermion] (v2) -- [fermion, edge label=\(q\)] (v), 
                (v1) -- [dashed, edge label=\(LQ\)] (v2) 
            };
          \end{feynman}
        \end{tikzpicture}
        \subcaption{Leptoquark contribution}
    \end{minipage}
    \caption{Representative one-loop BSM corrections that could induce a significant shift in $a_\tau$ compared to the SM value.}\label{fig:BSM}
\end{figure}
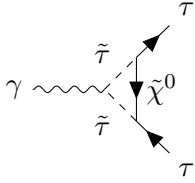
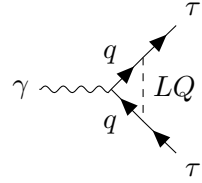

A key motivation for this hypothesis lies in the chirality-flip nature of the dipole operators. A common feature of many New Physics scenarios—including Supersymmetry, leptoquark models, and theories with additional gauge bosons—is that dipole operators are generated with a chirality flip proportional to the lepton mass. As a result, the correction to the anomalous magnetic moment scales with the square of the lepton mass, $\Delta a_\tau \propto (m_\tau/\Lambda_{NP})^2$, where $\Lambda_{NP}$ is the scale of New Physics. This mass-scaling law implies that the tau is approximately 280 times more sensitive than the muon to high-energy sectors, potentially allowing $a_\tau$ to reach values far beyond the SM prediction if new particles circulate in the radiative loops.
Among the most compelling scenarios are Supersymmetric models, where one-loop corrections involve sleptons ($\tilde{\tau}$) and neutralinos ($\tilde{\chi}^0$) \cite{Moroi:1995yh}. In these models, large values of $\tan\beta$ can significantly enhance the coupling, leading to a measurable shift in $a_\tau$. Similarly, the introduction of Leptoquarks (LQ) or new neutral gauge bosons ($Z'$) can generate sizable contributions depending on the model parameters. These BSM effects on $a_\tau$ are illustrated in the diagrams in Fig.~\ref{fig:BSM}.

The theoretical urgency for measuring $a_\tau$ is also deeply rooted in the persistent tension observed in the muon sector. Currently, the experimental world average for the muon magnetic anomaly, dominated by the Fermilab E989 results, shows a discrepancy with the Standard Model prediction. While the 2020 Theory Initiative White Paper (based on dispersive methods) points to a $5.1\sigma$ tension, recent Lattice QCD calculations—most notably by the BMW collaboration—suggest a larger hadronic vacuum polarization (HVP) contribution, which would significantly reduce this gap.

This ``HVP puzzle'' reinforces the importance of $a_\tau$ as an independent probe. In many BSM scenarios where New Physics couples to the lepton masses, the NP contributions are expected to scale as:
\begin{equation}
    \Delta a_\ell \propto \frac{m_\ell^2}{\Lambda_{NP}^2} \implies \frac{\Delta a_\tau}{\Delta a_\mu} \simeq \frac{m_\tau^2}{m_\mu^2} \approx 280
\end{equation}
Under this naive scaling~\footnote{This scaling behaviour typically arises in scenarios where the chirality flip occurs on the external lepton line.
However, it is important to note that this relation is not universal.
In models where the chirality flip is realized internally (e.g. through heavy fermion mixing or loop-induced mechanisms), or in the presence of non-minimal flavour structures, the scaling with $m_\ell^2$ can be significantly modified. Therefore, while the naive scaling $\Delta a_\ell \propto m_\ell^2$ provides a useful benchmark, it should not be regarded as a model-independent prediction.}, the muon anomaly ($\Delta a_\mu \sim 10^{-9}$) would translate into a shift $\Delta a_\tau \sim 10^{-7}$~\cite{Crivellin:2021spu}. 

Furthermore, in models involving Leptoquarks or $Z'$ bosons designed to explain $B$-physics flavor anomalies (see~\cite{DAlise:2024qmp, Vignaroli:2019lkg, Vignaroli:2018lpq} and references therein), the couplings to the third generation can be significantly enhanced beyond this scaling. In such cases, $\Delta a_\tau$ could reach values as high as $10^{-5}$ or $10^{-4}$, making it a decisive laboratory for distinguishing between different NP explanations of the muon's discrepancy.

When the energy scale $\Lambda$ of the New Physics affecting $a_\tau$ is significantly higher than the tau mass, its influence is most effectively captured using the Standard Model Effective Field Theory (SMEFT). This framework allows us to parameterize potential deviations in a model-independent way, translating experimental constraints on $a_\tau$ into bounds on the Wilson coefficients of higher-dimensional operators.

\subsection{EFT Parameterization}\label{sec:theory}

At energies below the electroweak scale $v$, the anomalous magnetic moment ($a_\tau$) and the electric dipole moment ($d_\tau$) can be parameterized through a dimension-5 effective Lagrangian that preserves $U(1)_{Q}$ gauge invariance:

\begin{equation}
\mathcal{L}_{eff}^{d=5} = -\frac{e}{4m_\tau} a_\tau (\bar{\tau} \sigma^{\mu\nu} \tau) F_{\mu\nu} + \frac{i}{2} d_\tau (\bar{\tau} \sigma^{\mu\nu} \gamma_5 \tau) F_{\mu\nu}
\end{equation}

While the low-energy properties of the tau lepton can be phenomenologically described by dimension-5 effective operators, a consistent theoretical framework must respect the full electroweak Standard Model gauge symmetry at energies ($\Lambda$) above the electroweak scale $v$. In the context of the Standard Model Effective Field Theory (SMEFT), dipole moments arise at leading order from two independent dimension-6 dipole operators 
that couple the leptonic fields to the Higgs doublet $\Phi$, preserving the SM $SU(2)_L\times U(1)_Y$ symmetry \cite{Grzadkowski:2010es, Brivio:2017vri}. The relevant terms in the SMEFT Lagrangian are given by:

\begin{equation}
\mathcal{L}_{\text{SMEFT}}^{d=6} = \frac{C_{ B \tau}}{\Lambda^2} (\bar{L}_\tau \sigma^{\mu\nu} \tau_R) \Phi B_{\mu\nu} + \frac{C_{  W \tau}}{\Lambda^2} (\bar{L}_\tau \sigma^{\mu\nu} \tau^I \tau_R) \Phi W_{\mu\nu}^I + \text{h.c.}
\end{equation}
where $L_\tau$ is the left-handed $SU(2)_L$ doublet (for the tauonic flavor), $\tau_R$ the right-handed singlet, $\tau^I$ are the Pauli matrices, and $B_{\mu\nu}, W_{\mu\nu}^I$ are the field strength tensors for $U(1)_Y$ and $SU(2)_L$, respectively.
After Spontaneous Symmetry Breaking (SSB), with $\langle \Phi \rangle = (0, v/\sqrt{2})^T$, the gauge fields ($B, W^I$) rotate into the physical basis ($A,Z,W^\pm$) giving rise to the following effective interactions:  
\begin{equation}
\begin{aligned}
\mathcal{L}_{int} &= \frac{v}{\sqrt{2}\Lambda^2} (\bar{\tau}_L \sigma^{\mu\nu} \tau_R) \left[ C_{ \gamma \tau} F_{\mu\nu} + C_{ Z \tau} Z_{\mu\nu} \right] \\
&+ \frac{v}{\Lambda^2} C_{ W \tau} \left[ (\bar{\nu}_\tau \sigma^{\mu\nu} \tau_R) W_{\mu\nu}^+ \right]  +\text{h.c.} + \dots
\end{aligned}\label{eq:Lint}
\end{equation}
where $F_{\mu\nu}$ and $Z_{\mu\nu}$ are, respectively, the photon and the $Z$ boson field strengths and $W^+_{\mu\nu} \equiv \partial_\mu W^+_\nu - \partial_\nu W^+_\mu$. The dipole couplings in the broken phase are matched to the Wilson coefficients in the unbroken phase via: 
\begin{equation}\label{eq:matching}
C_{\gamma \tau} = c_W C_{ B \tau} - s_W C_{ W \tau}, \quad C_{ Z \tau} = -s_W C_{ B \tau} - c_W C_{W \tau} 
\end{equation}
with $s_W (c_W)$ shortly denoting the sine (cosine) of the Weinberg angle.

By matching the effective dipole interaction onto the general vertex structure in Eq.~\eqref{eq:vertex}, one finds that the SMEFT contribution directly induces the Pauli form factor $F_2(0)$, leading to the relation:

\begin{equation}
a_\tau = \frac{4 m_\tau v}{\sqrt{2} e \Lambda^2} \text{Re}(C_{\gamma \tau}), \quad d_\tau = \frac{\sqrt{2} v}{\Lambda^2} \text{Im}(C_{\gamma \tau})
\end{equation}

This derivation highlights that $a_{\tau}$ is a direct probe of the real part of the Wilson coefficient of the photon dipole, while $d_{\tau}$ is related to the imaginary part. 

According to the matching relations in \eqref{eq:matching}, we see that, while $C_{B \tau}$ and $C_{W \tau}$ are independent coefficients, $C_{\gamma \tau}$ and $C_{Z \tau}$ are not independent.
The SMEFT framework reveals an intrinsic correlation between the photon and the $Z,W$-boson dipole moments.
Crucially, the $C_{Z\tau}$ term generates an anomalous $Z$-tau-tau dipole vertex, which implies that deviations in $a_\tau$ are intrinsically correlated to physics observable at the $Z$-pole. This correlation allows for powerful synergistic constraints between low-energy or UPC measurements and LEP/SLC data or future measurements at the FCC-ee~\cite{FCC:2018evy}. 
The term involving $W_{\mu\nu}^+$ demonstrates that the EFT framework intrinsically links the neutral dipole moments to charged current interactions $(\bar{\nu}_\tau \sigma^{\mu\nu} \tau_R) W_{\mu\nu}^+$, a correlation that can be tested in $W$ production channels. Furthermore, while we focus here on the single gauge boson vertex, the $SU(2)_L$ invariant dimension-6 operator also generates contact interactions (the dots in Eq.~\eqref{eq:Lint}) involving the tau lepton and two gauge bosons (e.g. $\tau\tau W W$, $\tau\nu W\gamma$ or $\tau\nu WZ$ ).  Specifically, these terms read:

\begin{equation}\label{eq:2bos-int}
\mathcal{L}^{2-bos}_{int} = i g \frac{v}{\Lambda^2} C_{ W \tau}  \,  (\bar{\nu}_\tau \sigma^{\mu\nu} \tau_R)  W^{+}_\mu (s_{W} A_\nu + c_{W} Z_\nu) + i g \frac{v}{\sqrt{2}\Lambda^2} C_{ W \tau}  \,(\bar{\tau}_L \sigma^{\mu\nu} \tau_R) \, W^{+}_\mu W^{-}_\nu  +\text{h.c.}
\end{equation}

Although these terms are negligible for current $a_\tau$ extractions,  these contributions can grow with energy and  could represent a critical frontier for futuristic high-energy experiments such as FCC-hh or a multi-TeV muon collider \cite{Vignaroli:2025pwn, Frigerio:2024jlh}.

This comprehensive gauge structure underscores why $a_\tau$ is not merely a precision constant, but a key to unlocking the full symmetry structure of New Physics.

\section{Experimental Probes at Particle Colliders}

The direct measurement of the tau anomalous magnetic moment $a_\tau$ is significantly more challenging than that for the electron or muon due to the tau's extremely short lifetime ($\approx 2.9 \times 10^{-13}\,\mathrm{s}$). Unlike muons, which can be stored in magnetic rings to observe their spin precession, tau leptons must be studied through scattering processes at high-energy colliders, where $a_\tau$ affects the production cross section and the angular distributions of the decay products. More specifically, sensitivity to $a_\tau$ arises through the contribution of the form factor $F_2(q^2)$. It is important to note that, in general, collider observables are sensitive to $F_2(q^2)$ evaluated at non-zero momentum transfer.
The extraction of the static quantity $a_\tau = F_2(0)$ therefore requires either an extrapolation to the limit $q^2 \to 0$ or an interpretation within a theoretical framework, such as SMEFT, relating measurements at different energy scales.
This distinction is crucial when comparing results obtained in different kinematic regimes.

\subsection{Historical Constraints from LEP: The DELPHI Measurement}
\begin{figure}[h]
    \centering
    \begin{tikzpicture}
      \begin{feynman}
        
        \vertex (a1) {\(e^-\)};
        \vertex [right=1.5cm of a1] (v1); 
        \vertex [right=1.5cm of v1] (b1) {\(e^-\)};
        
        \vertex [below=3.0cm of a1] (a2) {\(e^+\)};
        \vertex [right=1.5cm of a2] (v2); 
        \vertex [right=1.5cm of v2] (b2) {\(e^+\)};
        
        
        \vertex [below right=0.8cm and 1.2cm of v1, dot, fill=gray!80] (v3) {};
        \vertex [above right=0.8cm and 1.2cm of v2, dot, fill=gray!80] (v4) {};
        
        \vertex [right=2.0cm of v3] (f1) {\(\tau^-\)};
        \vertex [right=2.0cm of v4] (f2) {\(\tau^+\)};
    
        \diagram* {
          
          (a1) -- [fermion] (v1) -- [fermion] (b1),
          (b2) -- [fermion] (v2) -- [fermion] (a2), 
          
          (v1) -- [photon, edge label'=\(\gamma\)] (v3),
          (v2) -- [photon, edge label=\(\gamma\)] (v4),
          
          (v4) -- [fermion, edge label=\(\tau\)] (v3),
          
          (v3) -- [fermion] (f1),         
          (f2) -- [fermion] (v4),         
        };
      \end{feynman}
    \end{tikzpicture}
    \caption{Feynman diagram for the $e^+ e^-\to \tau^+\tau^- e^+e^-$ process at LEP via photon-photon fusion.  The blobs at the $\gamma\tau\tau$ vertices represent the effective interaction where the anomalous magnetic moment $a_\tau$ modifies the standard QED vertex.}    
    \label{fig:feynman_delphi}
\end{figure}
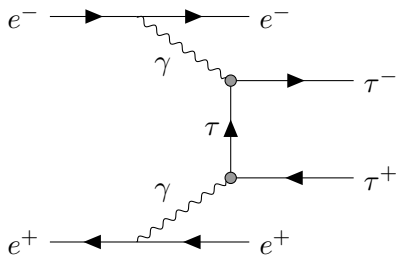

The longest-standing experimental bound on $a_\tau$ was established by the DELPHI collaboration at the Large Electron-Positron (LEP) collider \cite{Abdallah:2003xd}. The measurement utilized the $e^+ e^- \to e^+ e^- \tau^+ \tau^-$ process, where the tau pair is produced via photon-photon fusion ($\gamma\gamma \to \tau^+\tau^-$), as illustrated in Figure \ref{fig:feynman_delphi}.

In this framework, the presence of a non-zero $a_\tau$ modifies the electromagnetic vertex $\Gamma^\mu = \gamma^\mu + \frac{i\sigma^{\mu\nu}q_\nu}{2m_\tau}a_\tau$, leading to an enhancement of the total cross-section. The cross-section can be analytically parameterized as:
\begin{equation}
\sigma(a_\tau) = \sigma_{SM} + a_\tau \sigma_{1} + a_\tau^2 \sigma_{2}
\end{equation}
In this expansion, the linear term $\sigma_{1}$ represents the interference between the Standard Model tree-level amplitude and the amplitude contribution from the anomalous magnetic moment operator. The quadratic term $\sigma_{2}$ accounts for the pure anomalous magnetic contribution. By comparing the observed event yields with the SM expectation, DELPHI established the $95\%$ C.L. limit \cite{Abdallah:2003xd}:
\begin{equation}
-0.052 < a_\tau < 0.013
\end{equation}
While this result provided a fundamental constraint for nearly two decades, its precision ($\mathcal{O}(10^{-2})$) is still limited by the available luminosity and the clean but relatively low-energy environment of LEP compared to the LHC's runs.

\section{LHC Probes via Photon-Photon Fusion}
The experimental strategy for determining $a_\tau$ at high-energy colliders relies on measuring the cross-section and analyzing kinematic properties of the photon-photon fusion process, $\gamma\gamma \to \tau^+\tau^-$. At the LHC, this process is studied by exploiting the electromagnetic fields generated by the ultra-relativistic beams in two primary regimes: Proton-Proton ($pp$) collisions and Heavy-Ion Ultra-Peripheral Collisions (UPC).

\subsection{The Proton-Proton (pp) Channel}
\begin{figure}[h!]
    \centering
    \begin{tikzpicture}
      \begin{feynman}
        \vertex (a1) {\(p\)}; \vertex [right=1.2cm of a1] (v1); \vertex [right=1.2cm of v1] (b1) {\(p\)};
        \vertex [below=2.8cm of a1] (a2) {\(p\)}; \vertex [right=1.2cm of a2] (v2); \vertex [right=1.2cm of v2] (b2) {\(p\)};
        \vertex [below right=0.9cm and 0.8cm of v1, dot, fill=gray!80] (v3) {};
        \vertex [above right=0.9cm and 0.8cm of v2, dot, fill=gray!80] (v4) {};
        \vertex [right=2.2cm of v3] (f1) {\(\tau^-\)}; \vertex [right=2.2cm of v4] (f2) {\(\tau^+\)};
        \diagram* {
          (a1) -- (v1) -- (b1), (a2) -- (v2) -- (b2),
          (v1) -- [photon, edge label'=\(\gamma\)] (v3), (v2) -- [photon, edge label=\(\gamma\)] (v4),
          (v4) -- [fermion] (v3), (v3) -- [fermion] (f1), (f2) -- [fermion] (v4)
        };
        \node [below=0.2cm of v2] {\footnotesize \textbf{(a) Elastic}};

        \begin{scope}[xshift=5.8cm]
        \vertex (sa1) {\(p\)}; \vertex [right=1.2cm of sa1] (sv1); \vertex [right=1.2cm of sv1] (sb1) {\(p\)};
        \vertex [below=2.8cm of sa1] (sa2) {\(p\)}; \vertex [right=1.2cm of sa2] (sv2);
        \vertex [right=1.2cm of sv2] (Xc) {\(X\)};
        \vertex [above right=0.4cm and 1.3cm of sv2] (Xu); \vertex [below right=0.4cm and 1.3cm of sv2] (Xd);
        \vertex [below right=0.9cm and 0.8cm of sv1, dot, fill=gray!80] (sv3) {};
        \vertex [above right=0.9cm and 0.8cm of sv2, dot, fill=gray!80] (sv4) {};
        \vertex [right=2.2cm of sv3] (sf1) {\(\tau^-\)}; \vertex [right=2.2cm of sv4] (sf2) {\(\tau^+\)};
        \diagram* {
          (sa1) -- (sv1) -- (sb1), (sa2) -- (sv2), 
          (sv2) -- [dashed] (Xc), (sv2) -- [dashed] (Xu), (sv2) -- [dashed] (Xd),
          (sv1) -- [photon, edge label'=\(\gamma\)] (sv3), (sv2) -- [photon, edge label=\(\gamma\)] (sv4),
          (sv4) -- [fermion] (sv3), (sv3) -- [fermion] (sf1), (sf2) -- [fermion] (sv4)
        };
        \node [below=0.2cm of sv2] {\footnotesize \textbf{(b) Semi-elastic}};
        \end{scope}

        \begin{scope}[xshift=11.6cm]
        \vertex (ia1) {\(p\)}; \vertex [right=1.2cm of ia1] (iv1);
        \vertex [right=1.2cm of iv1] (iXc) {\(X\)};
        \vertex [above right=0.4cm and 1.3cm of iv1] (iXu); \vertex [below right=0.4cm and 1.3cm of iv1] (iXd);
        \vertex [below=2.8cm of ia1] (ia2) {\(p\)}; \vertex [right=1.2cm of ia2] (iv2);
        \vertex [right=1.2cm of iv2] (iYc) {\(X'\)};
        \vertex [above right=0.4cm and 1.3cm of iv2] (iYu); \vertex [below right=0.4cm and 1.3cm of iv2] (iYd);
        \vertex [below right=0.9cm and 0.8cm of iv1, dot, fill=gray!80] (iv3) {};
        \vertex [above right=0.9cm and 0.8cm of iv2, dot, fill=gray!80] (iv4) {};
        \vertex [right=2.2cm of iv3] (if1) {\(\tau^-\)}; \vertex [right=2.2cm of iv4] (if2) {\(\tau^+\)};
        \diagram* {
          (ia1) -- (iv1), (iv1) -- [dashed] (iXc), (iv1) -- [dashed] (iXu), (iv1) -- [dashed] (iXd),
          (ia2) -- [fermion] (iv2), (iv2) -- [dashed] (iYc), (iv2) -- [dashed] (iYu), (iv2) -- [dashed] (iYd),
          (iv1) -- [photon, edge label'=\(\gamma\)] (iv3), (iv2) -- [photon, edge label=\(\gamma\)] (iv4),
          (iv4) -- [fermion] (iv3), (iv3) -- [fermion] (if1), (if2) -- [fermion] (iv4)
        };
        \node [below=0.2cm of iv2] {\footnotesize \textbf{(c) Inelastic}};
        \end{scope}
      \end{feynman}
    \end{tikzpicture}
    \caption{Feynman diagrams for $\tau^+\tau^-$ production in $pp$ collisions. The gray blobs represent the $a_\tau$-sensitive effective vertices.}
    \label{fig:pp_channels}
\end{figure}
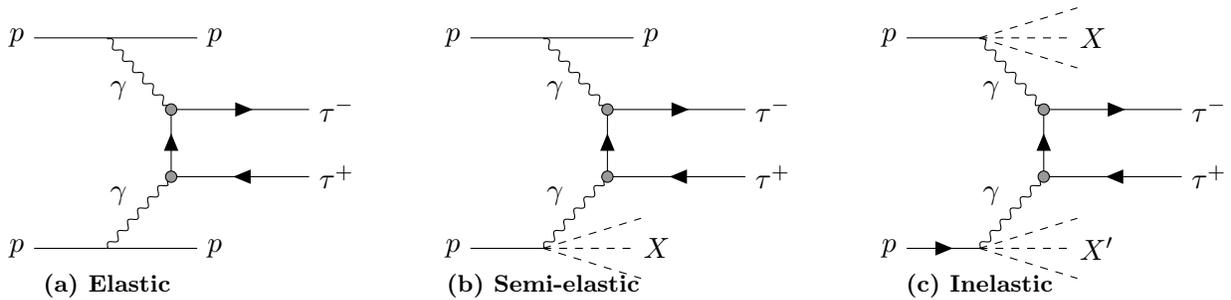

In $pp$ collisions, the production of tau pairs via photon exchange is a rare but powerful probe. The primary challenge compared to the LEP environment arises from the composite nature of the proton. As illustrated in Figure \ref{fig:pp_channels}, three distinct interaction topologies are defined by the momentum transfer and the resulting integrity of the final-state protons:

\begin{itemize}
    \item \textbf{Elastic:} Both protons emit a quasi-real photon and remain intact, escaping the interaction point through the beam pipe. Experimentally, these events are uniquely clean, characterized by a rapidity gap (a lack of hadronic activity surrounding the tau production vertex).
    \item \textbf{Semi-elastic (Single-Dissociative):} One proton remains intact while the other, due to the photon emission's recoil, dissociates into a low-mass hadronic system $X$. 
    \item \textbf{Inelastic (Double-Dissociative):} Both protons dissociate into hadronic fragments. While this channel provides the highest raw statistics, it introduces significant systematic uncertainties related to the modeling of the proton's internal structure (photon PDFs) and the ``survival probability'' of the central exclusive signature.
\end{itemize}

Recent observations by CMS (2024) \cite{CMS:2024qjo} in this channel 
has demonstrated that despite the challenging background environment, $pp$ data can yield very competitive constraints on $a_\tau$. This analysis relies on advanced track-counting algorithms to suppress hadronic backgrounds and isolate the electroweak signal, effectively turning the LHC into a high-energy photon-photon collider.

\subsection{The Ultra-Peripheral Collision (UPC) Channel}
Ultra-Peripheral Collisions (UPC) of heavy ions, specifically Lead ($^{208}$Pb), provide a particularly clean and theoretically controlled environment at the LHC for probing the tau anomalous magnetic moment. These interactions occur when the impact parameter $b$ exceeds the sum of the nuclear radii ($b > R_1 + R_2$), ensuring that the short-range strong force is suppressed and the interaction is mediated by the coherent exchange of quasi-real photons.

\begin{figure}[h!]
    \centering
    \begin{tikzpicture}[scale=1.3, every node/.style={transform shape}]
      \begin{feynman}
        \vertex (a1) {\(Pb\)}; \vertex [right=1.5cm of a1] (v1); \vertex [right=1.5cm of v1] (b1) {\(Pb\)};
        \vertex [below=3cm of a1] (a2) {\(Pb\)}; \vertex [right=1.5cm of a2] (v2); \vertex [right=1.5cm of v2] (b2) {\(Pb\)};
        \vertex [below right=0.8cm and 0.8cm of v1, dot, fill=gray!80] (v3) {};
        \vertex [above right=0.8cm and 0.8cm of v2, dot, fill=gray!80] (v4) {};
        \vertex [right=2.2cm of v3] (f1) {\(\tau^-\)}; \vertex [right=2.2cm of v4] (f2) {\(\tau^+\)};
        \diagram* {
          (a1) -- [double, double distance=2pt] (v1) -- [double, double distance=2pt] (b1),
          (a2) -- [double, double distance=2pt] (v2) -- [double, double distance=2pt] (b2),
          (v1) -- [photon, edge label'=\(\gamma\)] (v3), (v2) -- [photon, edge label=\(\gamma\)] (v4),
          (v4) -- [fermion, edge label=\(\tau\)] (v3), (v3) -- [fermion] (f1), (f2) -- [fermion] (v4)
        };
      \end{feynman}
    \end{tikzpicture}
    \caption{Photon-photon fusion in $PbPb$ UPC. The $Z^4$ enhancement of the cross-section allows for a high-precision study of the $\gamma\tau\tau$ vertex.}
    \label{fig:upc_fermi}
\end{figure}
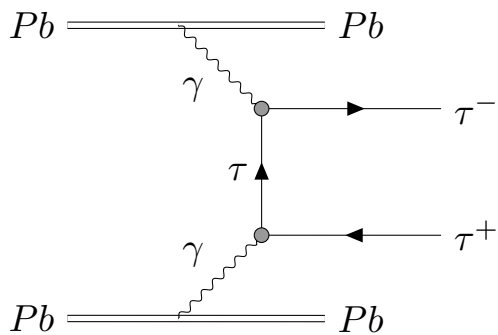

The theoretical foundation for calculating the production of tau pairs in UPC is the Method of Equivalent Photons, originally proposed by Fermi in 1924 \cite{Fermi:1924tc} and later refined by Weizsäcker and Williams \cite{vonWeizsacker:1934nji, Williams:1934ad}. In this framework, the electromagnetic fields of ultra-relativistic ions are treated as a beam of nearly on-shell photons.  
The leading-order process for this interaction, where two photons from the colliding nuclei fuse into a tau pair ($\gamma\gamma \to \tau^+\tau^-$), is illustrated in Fig. \ref{fig:upc_fermi}.

The total cross-section for the process $PbPb \to PbPb \tau^+\tau^-$ is expressed as a convolution of the photon fluxes from the two colliding nuclei:

\begin{equation}\label{eq:cross-section}
    \sigma(PbPb \to PbPb\tau^+\tau^-) = \int dx_1 dx_2 \, n(x_1) n(x_2) \, \sigma_{\gamma\gamma \to \tau^+\tau^-}(W_{\gamma\gamma})
\end{equation}

where $x_i = \omega_i / E_{beam}$ is the fraction of the beam energy carried by the photon, $n(x)$ is the equivalent photon flux, and $\sigma_{\gamma\gamma \to \tau^+\tau^-}$ is the elementary cross-section for the fusion process at the invariant mass $W_{\gamma\gamma} = \sqrt{x_1 x_2 s}$.

The photon flux $n(x)$ represents the intensity of the electromagnetic field surrounding the lead nucleus ($Z=82$). 
While the general expression depends on the nuclear form factor $F(q^2)$ \cite{Budnev:1975poe, Vidovic:1992ik},  for a relativistic nucleus the flux can be expressed in terms of modified Bessel functions of the second kind of the first and second order ($K_0, K_1$) for a point-like charge approximation at a given impact parameter $b$:

\begin{equation}
    N(\omega, b) = \frac{Z^2 \alpha}{\pi^2 \beta^2 b^2} \left[ \xi^2 K_1^2(\xi) + \frac{\xi^2}{\gamma_L^2} K_0^2(\xi) \right]
\end{equation}

where $\xi = \omega b / \gamma_L \beta \hbar c$ is a dimensionless variable and $\gamma_L$ represents the Lorentz factor of the nucleus. The integrated flux $n(x)$ is calculated by imposing the exclusivity condition $b > 2R$ to avoid hadronic overlap:

\begin{equation}
    n(x) = \int_{2R}^{\infty} 2\pi b \, db \, N(x E_{beam}, b)
\end{equation}

By performing the integration over the impact parameter $b$ from $R_{min} = 2R$ to infinity, the equivalent photon flux $n(x)$ can be expressed in the closed analytical form typically employed in recent LHC studies \cite{Beresford:2019gww, Verducci:2023cgx}:

\begin{equation}
    n(x) = \frac{2 Z^2 \alpha}{\pi \beta^2 x} \left[ \zeta K_0(\zeta) K_1(\zeta) - \frac{\beta^2 \zeta^2}{2} \left( K_1^2(\zeta) - K_0^2(\zeta) \right) \right]
\end{equation}

where the dimensionless scaling variable is now defined as $\zeta = x E_{beam} R_{min} / \gamma_L \beta \hbar c$. This expression properly accounts for the finite size of the nuclei through the cutoff $R_{min}$, ensuring that the measurement remains in the ultra-peripheral regime. 
The cross section in Eq.~\eqref{eq:cross-section} can be also expressed in terms of an effective $\gamma\gamma$ luminosity $(\frac{dL_{\text{eff}}}{dW_{\gamma\gamma}})$ as

\begin{equation}\label{eq:xsec-eff-lum}
 \sigma(PbPb \to PbPb\tau^+\tau^-) = \int dW_{\gamma\gamma} \frac{dL_{\text{eff}}}{dW_{\gamma\gamma}} \sigma_{\gamma\gamma \to \tau^+\tau^-}(W_{\gamma\gamma}) \, .
\end{equation}

This treatment highlights the $Z^4$ enhancement of the total cross-section, which provides a significant enhancement of the signal rate, while the requirement of ``exclusivity'' ensures the ``clean'' environment necessary to isolate the $a_\tau$ contribution from hadronic backgrounds.
An important advantage of the UPC framework is the relative suppression of nuclear effects. As discussed in  \cite{Beresford:2019gww, Dyndal:2020yen, Verducci:2023cgx}, for the kinematic regions relevant to $\tau$-pair production, the impact of nuclear shadowing and the transition from coherent to incoherent photon emission are well-constrained. Specifically, the large impact parameters required for ultra-peripheral events ensure that hadronic interactions are negligible, while the coherent nature of the photon flux, which scales with $Z^2$, remains the dominant and theoretically clean production mechanism. This allows for the treatment of the lead ions as quasi-real photon sources, with nuclear form factor uncertainties contributing only marginally to the overall systematic budget.

Transformation of the LHC into a high-intensity photon collider to probe $a_\tau$ was first proposed in \cite{Beresford:2019gww} and \cite{Dyndal:2020yen}. These studies identified Ultra-Peripheral Collisions as the ideal environment to break the precision barriers set by LEP, highlighting the sensitivity of the $\gamma\gamma \to \tau\tau$ cross-section to $a_\tau$ at the LHC energies.

On the experimental front, the ATLAS Collaboration in 2022 provided the first measurement of $a_\tau$ in this channel using Run 2 $PbPb$ data, observing the process with a significance of $5.0\sigma$~ \cite{ATLAS:2022ryk}. This milestone demonstrated that the UPC environment allows for a significant suppression of backgrounds, yielding a significant constraint on $a_\tau$. 
This was followed by the CMS Collaboration~\cite{CMS:2022arf}, which utilized similar UPC techniques and further refined the measurement, approaching the sensitivity of the combined LEP results.
 The current 95\% CL exclusion limits for these experiments are as follows:

\begin{equation}
    \text{ATLAS (2022): } a_\tau \in [-0.057, 0.024] \quad ; \quad \text{CMS (2022): } a_\tau \in [-0.030, 0.017]
\end{equation}

The current research landscape is characterized by several key developments:
\begin{itemize}
    \item \textbf{Advanced Analysis Strategies:} Recent studies, such as \cite{Verducci:2023cgx}, focus on optimizing the experimental selection through Multivariate Analysis (MVA) and BDT-based discrimination to separate the $\tau\tau$ signal from the $\mu\mu$ background.
    \item \textbf{Higher-Order Corrections:} To match the increasing experimental precision, theoretical efforts have addressed electroweak and QED corrections to the UPC $\tau^+\tau^-$ production \cite{Dittmaier:2025ikh}.
    \item \textbf{Quantum Observables:} New frontier studies suggest probing $a_\tau$ and $d_\tau$ through quantum tomography and spin correlations in the final state \cite{LoChiatto:2024dmx, Shao:2023bga}.
\end{itemize}

Looking forward, the integration of larger data sets and the potential use of forward proton detectors in $pp$ collisions \cite{Beresford:2024dsc} will provide a complementary dataset to the $PbPb$ UPC measurements, as discussed in the following section.

\section{Complementarity between $pp$ and UPC Channels}

\begin{figure}[h!]
    \centering
    \includegraphics[width=0.6\textwidth]{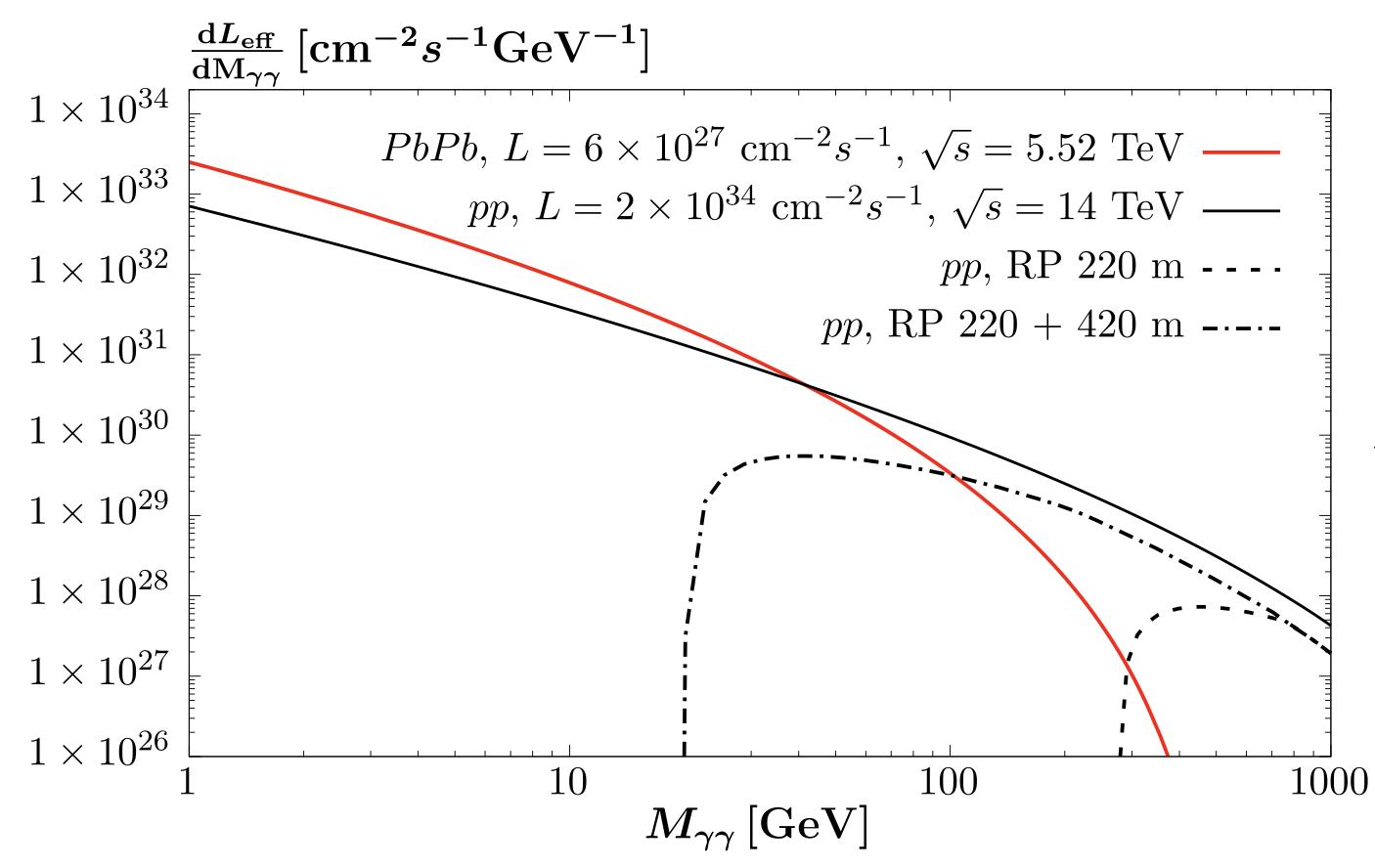}
    \caption{Effective photon-photon luminosity as a function of the invariant mass $M_{\gamma\gamma}\equiv W_{\gamma\gamma}$ as defined in Eq.~\eqref{eq:xsec-eff-lum}. The plot highlights the $Z^4$ dominance of $PbPb$ at low masses, where the measurement of $a_\tau$ is more direct, and the superior reach of $pp$ collisions towards the TeV energy frontier. Figure adapted from \cite{Bruce:2018yzs}.}
    \label{fig:luminosity_comparison}
\end{figure}

The investigation of $a_\tau$ at the LHC relies on a strategic synergy between proton-proton ($pp$) and heavy-ion Ultra-Peripheral Collisions ($PbPb$ UPC). While both channels exploit the $\gamma\gamma \to \tau^+\tau^-$ process, they probe fundamentally different kinematic regimes due to the disparate spatial scales and luminosities of the colliding species, as illustrated in Fig.~\ref{fig:luminosity_comparison}.

The complementarity of these two environments can be summarized across these main axes:

\begin{itemize}
    \item \textbf{Energy Reach and Nuclear Radius:} The maximum energy of the equivalent photons is inversely proportional to the radius $R$ of the charge distribution ($\omega_{max} \approx \gamma_L \hbar c / R$). The large radius of the Lead nucleus ($R_{Pb} \approx 7$ fm) restricts UPCs to $W_{\gamma\gamma} \lesssim 100$ GeV, while the small proton radius ($R_{p} \approx 0.8$ fm) allows $pp$ collisions to reach the TeV scale.
    
    \item \textbf{Directness of the $a_\tau$ Measurement:} By definition, the anomalous magnetic moment is a static property derived from the effective $\gamma\tau\tau$ vertex in the limit of zero photon momentum ($q^2 \to 0$). Since UPCs occur predominantly near the kinematic threshold ($W_{\gamma\gamma} \approx 2m_\tau$), they provide a more \textit{direct} measurement of $a_\tau$ that is less dependent on the energy-running of the coupling. In contrast, $pp$ measurements at high $W_{\gamma\gamma}$ probe the dipole form factor at large momentum transfer, requiring an interpretation within an EFT framework to relate high-energy measurements to the static limit. 
    
    \item \textbf{Statistics and Integrated Luminosity:} Although $PbPb$ collisions benefit from the $Z^4$ enhancement of the photon flux, they are limited by the lower integrated luminosity delivered during heavy-ion runs (nb$^{-1}$ scale). $pp$ collisions compensate for the lack of $Z$-scaling by leveraging the full high-intensity phase of the LHC (100 fb$^{-1}$ scale), resulting in a significantly larger total dataset for tau pair production.
    
    \item \textbf{Experimental Environment and Systematics:} UPCs offer an exceptionally clean environment with almost zero pile-up, where the primary challenge is the modeling of the nuclear form factor and photon flux uncertainties. Conversely, $pp$ collisions are characterized by high pile-up, requiring advanced experimental techniques such as track-counting and vertex isolation to identify the exclusive signal. 
    \end{itemize}

From a theoretical perspective, the different experimental strategies do not probe a single observable, but rather different kinematic regimes of the same underlying form factor $F_2(q^2)$.
UPCs access the near-static limit, while high-energy $pp$ collisions probe the ultraviolet behaviour of the effective interaction.
This complementarity is therefore not only experimental, but intrinsically linked to the momentum dependence of the electromagnetic vertex.

This dual approach ensures that any potential deviation from the Standard Model observed in one channel can be validated in a completely different kinematic and experimental environment. However, the interpretation of high-energy $pp$ data requires a careful assessment of the theoretical framework, particularly regarding the validity of the EFT expansion as the collision energy approaches the scale of New Physics.

\section{Recent Experimental Results and Outlook}

The experimental determination of the tau anomalous magnetic moment ($a_\tau$) has recently undergone a paradigm shift. While traditionally confined to low-energy $e^+e^-$ colliders, the LHC has emerged as a powerful tool for probing the $\gamma\tau\tau$ vertex across diverse kinematic regimes. The current experimental landscape, summarized in Table \ref{tab:results_summary}, highlights a clear distinction between measurements performed in the quasi-static limit and those probing the vertex at high virtualities.

\begin{table}[h]
\centering
\caption{Current 95\% CL constraints on $a_\tau$. The Standard Model prediction is $a_\tau^{\text{SM}} \approx 117\,721 \times 10^{-8}$. For $PbPb$ collisions, $\sqrt{s_{NN}}$ denotes the energy per nucleon pair.}
\label{tab:results_summary}
\begin{small}
\begin{tabular}{lcccc}
\hline
\textbf{Experiment} & \textbf{Channel} & \textbf{$\sqrt{s}$ or $\sqrt{s_{NN}}$} & \textbf{Dataset} & \textbf{Exclusion Limit (95\% CL)} \\ \hline
DELPHI (LEP) & $e^+e^- \to e^+e^-\tau^+\tau^-$ & $183 - 208$ GeV & 650 pb$^{-1}$ & $[-0.052, 0.013]$ \\
ATLAS (2022) & $PbPb$ UPC & 5.02 TeV & 1.44 nb$^{-1}$ & $[-0.057, 0.024]$ \\
CMS (2022)   & $PbPb$ UPC & 5.02 TeV & 1.34 nb$^{-1}$ & $[-0.030, 0.017]$ \\
CMS (2024)   & $pp$ Exclusive  \cite{CMS:2024qjo}* & 13 TeV & 138 fb$^{-1}$ & $[-0.0022, 0.0041]$ \\
ATLAS (2025) & $pp$ High-Mass  \cite{ATLAS:2025oiy}** & 13 TeV & 140 fb$^{-1}$ & $[-0.0024, 0.0047]$ \\ \hline
\end{tabular}
\end{small}
\end{table}
\textit{*Measurement performed in an intermediate energy regime ($W_{\gamma\gamma} \in [50, 500]$ GeV).} \\
\textit{**Value derived within the SMEFT framework at a scale $\Lambda = 1$ TeV.}

\paragraph{Interpretation of high-energy constraints.}

It is important to stress that, at high energies, collider measurements do not directly probe the static anomalous magnetic moment $a_\tau = F_2(0)$.
Instead, they are sensitive to the momentum-dependent form factor $F_2(q^2)$ at large momentum transfer, $q^2 \sim \hat{s}$.

In this regime, experimental constraints are more rigorously interpreted as bounds on the Wilson coefficients of higher-dimensional operators within the SMEFT framework.
The commonly quoted limits on $a_\tau$ are therefore derived quantities, obtained under the assumption of EFT validity and truncation at dimension-6.

Strictly speaking, these measurements constrain combinations of operator coefficients rather than a single static observable.

\subsection{From Quasi-Static Probes to High-Energy $pp$ Collisions}

Before discussing the latest results in proton-proton ($pp$) collisions, it is essential to consider the role of Ultra-Peripheral Collisions (UPC) in heavy-ion ($PbPb$) runs. As shown in Table \ref{tab:kinematics}, these measurements are performed in a kinematic regime where the photon virtuality is extremely low ($q^2 \approx 0$). Consequently, UPCs provide the most ``robust'' measurements of $a_\tau$ in the sense that they are the closest experimental equivalent to a static limit probe at a hadron collider. Although UPCs probe very small photon virtualities, ,corresponding to a near-threshold configuration for $\tau$-pair production, they do not correspond exactly to the static limit, but rather to a regime where the momentum transfer is minimal and theoretically well controlled.
By minimizing the energy transfer to the $\tau\tau$ system ($W_{\gamma\gamma} < 50$ GeV), these analyses remain largely independent of the complexities associated with High-Dimension operators in the Effective Field Theory (EFT) expansion.

Building upon this foundation, the CMS Collaboration \cite{CMS:2024qjo} recently observed the $\gamma\gamma \to \tau\tau$ process in $pp$ collisions for the first time. This analysis leverages the full Run 2 integrated luminosity (138~fb$^{-1}$) to isolate tau pairs produced via photon fusion. As illustrated in Fig. \ref{fig:pp_channels}, the signal includes contributions from elastic, semi-elastic, and inelastic proton scattering. To separate these events from the overwhelming Drell-Yan background, CMS employs a selection based on the acoplanarity of the tau decay products and the absence of charged tracks at the production vertex. This allows the measurement to probe an intermediate kinematic window ($W_{\gamma\gamma} \in [50, 500]$ GeV) where the EFT expansion is expected to remain under control, provided the relevant energy scales remain below the cutoff ($W \ll \Lambda$), yielding a limit nearly an order of magnitude more precise than LEP.

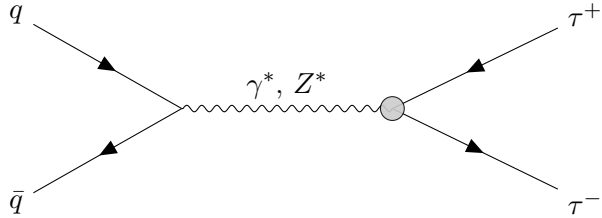
\begin{figure}[h!]
\centering
\begin{tikzpicture}
  \begin{feynman}
    \vertex (i1) {\(q\)};
    \vertex [below=2.5cm of i1] (i2) {\(\bar{q}\)};
    \vertex [right=2.2cm of i1, yshift=-1.25cm] (v1);
    \vertex [right=2.8cm of v1] (v2);
    \vertex [right=2.2cm of v2, yshift=1.25cm] (f1) {\(\tau^{+}\)};
    \vertex [right=2.2cm of v2, yshift=-1.25cm] (f2) {\(\tau^{-}\)};
    \diagram* {
      (i1) -- [fermion] (v1),
      (v1) -- [fermion] (i2), 
      (v1) -- [photon, edge label=\(\gamma^{*}\text{, } Z^{*}\)] (v2),
      (f1) -- [fermion] (v2), 
      (v2) -- [fermion] (f2),
    };
    \filldraw[fill=gray!40, draw=black, opacity=0.9] (v2) circle (4.5pt);
  \end{feynman}
\end{tikzpicture}
\caption{Feynman diagram for the Drell-Yan process $q\bar{q} \to \tau^+\tau^-$ in $pp$ collisions. The grey blob represents the effective vertex modification probed by ATLAS.}
\label{fig:drell_yan}
\end{figure}

A complementary approach is provided by the ATLAS Collaboration \cite{ATLAS:2025oiy}, which investigates the high-mass tail of the inclusive Drell-Yan process ($q\bar{q} \to \tau^+\tau^-$). Unlike photon-fusion, this channel probes the $\tau\tau V$ vertex through $s$-channel exchange of virtual photons and $Z$ bosons (see Fig. \ref{fig:drell_yan}).

The sensitivity of this analysis relies on the energy-dependent enhancement of SMEFT operators, where the BSM contribution scales as $s/\Lambda^2$. ATLAS sets limits on the Wilson coefficient $C_{\gamma\tau}/\Lambda^2$, which is matched to $a_\tau$ via the relation:
\begin{equation}
  a_\tau = \frac{4 m_\tau v}{\sqrt{2} e \Lambda^2} \text{Re}(C_{\gamma \tau})
\end{equation}
For $\Lambda = 1$ TeV, ATLAS obtains $a_\tau \in [-0.0024, 0.0047]$. 

However, as summarized in Table \ref{tab:kinematics}, the interpretation approaches the regime where the validity of the EFT expansion may become questionable: the sensitivity is driven by events where $\sqrt{\hat{s}}$ reaches the TeV scale, challenging the assumption of scale separation ($\sqrt{\hat{s}} \ll \Lambda$). At these energies, the $a_\tau$ contribution is inextricably linked to both $\gamma^*$ and $Z^*$ exchanges. 

\begin{table}[h!]
\centering
\caption{Kinematic regimes and theoretical robustness of LHC $a_\tau$ probes.}
\label{tab:kinematics}
\begin{tabular}{lccc}
\hline
\textbf{Channel} & \textbf{Typical $W_{\gamma\gamma}$ or $\sqrt{\hat{s}}$} & \textbf{Photon State} & \textbf{EFT Validity} \\ \hline
$PbPb$ UPC & $< 50$ GeV & Quasi-real ($q^2 \approx 0$) & Robust (Static) \\
$pp$ Exclusive (CMS) & $50 - 500$ GeV & Moderate virtuality & Safe ($W \ll \Lambda$) \\
$pp$ High-Mass (ATLAS) & $500 - 2000$ GeV & Highly virtual & Borderline ($W \approx \Lambda$) \\ \hline
\end{tabular}
\end{table}

A crucial subtlety indeed arises in the interpretation of high-energy Drell--Yan measurements. 
While the anomalous magnetic moment $a_\tau$ is defined in the static limit $q^2 \to 0$, 
the processes probed at the LHC involve momentum transfers in the TeV range, $q^2 \sim \hat{s} \gg m_\tau^2$. 

In this regime, the SMEFT dipole operators induce a momentum-dependent modification 
of the $\gamma^\ast \tau\tau$ and $Z^\ast \tau\tau$ vertices, effectively probing the form factor $F_2(q^2)$ away from the static limit. 
As a result, the extracted constraints do not correspond to a direct measurement of $a_\tau$, 
but rather to bounds on the Wilson coefficient $C_{\gamma \tau}/\Lambda^2$ under the assumption of EFT validity.

Moreover, at high energies the dipole contribution interferes with both photon- and $Z$-mediated amplitudes, 
and can be accompanied by additional contact interactions arising from the same dimension-6 operators. 
This implies that the interpretation in terms of a single parameter $a_\tau$ becomes model-dependent, 
especially in the regime $\sqrt{\hat{s}} \sim \Lambda$. 
In this regime, the validity of the EFT expansion becomes non-trivial.
When the typical energy scale of the process approaches the cutoff, higher-dimensional operators (e.g. dimension-8) may give non-negligible contributions.

Moreover, the truncation of the EFT expansion at dimension-6 may lead to inconsistencies, including potential violations of perturbative unitarity.
For this reason, a consistent EFT interpretation requires either restricting the analysis to $\sqrt{\hat{s}} \ll \Lambda$ or adopting more refined approaches, such as including higher-order operators or performing validity-driven event selections.

\subsection{Future Facilities and Projections}

The quest for $a_{\tau}$ is set to advance along two complementary paths: the \textit{precision frontier}, dominated by $e^+ e^-$ lepton colliders, and the \textit{energy frontier}, explored at hadron and ion facilities and at a multi-TeV muon collider. The analysis of recent projections reveals a highly nuanced landscape where sensitivities vary by several orders of magnitude.

\paragraph{The Precision Frontier: Belle II and FCC-ee} 

Lepton colliders offer the cleanest environment for $a_{\tau}$ determination. 
\begin{itemize}
    \item \textbf{Belle II (SuperKEKB):} 
    Belle II exploits the massive production of $\tau^+\tau^-$ pairs from $e^+e^-$ collisions at the $\Upsilon(4\mathrm{S})$ resonance ($\sqrt{s} = 10.58$~GeV). Current analyses, such as those in \cite{Bodrov:2024wrw}, leverage spin correlations in hadronic decays (e.g., $\tau^\pm \to \pi^\pm \nu_\tau$), where the decay products act as effective spin polarimeters. While these measurements currently set the most stringent direct limits, the ultimate sensitivity is expected from the proposed upgrade featuring a longitudinally polarized electron beam. As demonstrated in \cite{Bernabeu:2008ii}, measuring left-right asymmetries ($A_{LR}$) would allow for a precision of $\mathcal{O}(10^{-5})$. This would potentially enable the first observation of the Standard Model value and provide sensitivity to New Physics contributions, provided that systematic uncertainties are controlled below the 0.5\% level. However, reaching such high precision requires a rigorous control of subleading theoretical effects. Recent work \cite{Gogniat:2026amk} highlights that at the Belle II energy scale, electroweak contributions from $Z$-boson exchange arise at the level of $\simeq 3 \times 10^{-6}$. Furthermore, the potential impact of four-fermion operators—which could parameterize BSM scenarios other than dipole moments—needs to be carefully disentangled, as they could induce effects up to $\mathcal{O}(10^{-5})$. This underscores the necessity of a global approach to accurately isolate $a_{\tau}$ from other competing effective operators and radiative corrections.
    
    \item \textbf{FCC-ee:} Operating at the $Z$ pole, FCC-ee will produce approximately $1.3 \times 10^{11}$ $\tau$-pairs. Recent global analyses \cite{Buttazzo:2026amk, Dam:2018rfz} indicate that by utilizing $\tau$-polarization observables and analyzing the $e^+e^- \to \tau^+\tau^-\gamma$ radiative process, FCC-ee is expected to reach a precision of $\mathcal{O}(10^{-5})$. This level of sensitivity would allow for a model-independent test of the SM electroweak loop corrections for the first time.
\end{itemize}

\paragraph{The Energy Frontier: FCC-hh and High-Energy Lepton Colliders}
While high-energy environments provide high photon fluxes, their sensitivity to $a_{\tau}$ is notably different:
\begin{itemize}
    \item \textbf{FCC-hh (PbPb UPC):} Detailed studies of Ultra-Peripheral Collisions at $\sqrt{s_{NN}} = 39 \text{ TeV}$ show that, despite the $Z^4$ enhancement of the photon flux, the 95\% C.L. exclusion limits on the anomalous magnetic moment are projected at $|a_{\tau}| \le 0.0102$ \cite{Inan:2026byu}. This sensitivity, in the order of $10^{-2}$, is significantly less competitive than the lepton collider projections, though it provides a theoretically robust and independent test.
    \item \textbf{Muon Collider:} On a longer time horizon, the high-energy Muon Collider is emerging as a high-interest R\&D project~\cite{InternationalMuonCollider:2025sys}. As highlighted in \cite{Buttazzo:2026amk, Wang:2024bfc, Denizli:2024uwv}, a facility operating at 3--10 TeV could leverage Drell-Yan $\mu^+\mu^- \to \tau^+\tau^-(\gamma)$ and Vector-Boson-Fusion (VBF) processes to reach sensitivities of $\mathcal{O}(10^{-5}-10^{-4})$. Notably, the study of the radiative Higgs decay $h \to \tau\tau\gamma$ at a Muon Collider offers a novel and powerful probe, potentially reaching a resolution of $\mathcal{O}(10^{-6})$  \cite{Buttazzo:2026amk}, which would represent the ultimate frontier in testing the electromagnetic properties of the third-generation leptons.
\end{itemize}

In this context, it is important to highlight that at high energies, as previously demonstrated for neutrino dipole interactions \cite{Vignaroli:2025pwn, Frigerio:2024jlh}, the effective interactions involving two gauge bosons (Eq.~\eqref{eq:2bos-int}) generated by $SU(2)_L$-invariant operators—which are typically neglected in low-energy or near-threshold studies—offer a powerful and highly effective way to probe magnetic dipole moments through VBF channels. This approach, adopted for the tau lepton \cite{Vignaroli:in-preparation}, could significantly enhance the sensitivity of both high-energy muon colliders and the FCC-hh to $a_{\tau}$, effectively exploiting the multi-TeV reach of these facilities.

\section{Conclusions}

The anomalous magnetic moment of the tau lepton represents a unique probe of the Standard Model and its possible extensions in the third lepton generation. Unlike the electron and muon, its extremely short lifetime prevents direct measurements, making collider-based approaches essential.

A key message emerging from this review is the complementarity between different experimental strategies. 
The experimental landscape has recently undergone a major transformation, breaking the long-standing limits set by LEP. A key driver of this progress has been the exploitation of Ultra-Peripheral Collisions (UPCs) at the LHC. By acting as high-intensity photon-photon colliders, and thanks to the $Z^4$ enhancement of the photon flux in $PbPb$ runs, UPCs have provided among the most theoretically robust and clean environments, operating close to the static limit \cite{ATLAS:2022ryk,CMS:2022arf}. In parallel, new techniques in $pp$ collisions—such as the CMS measurement in exclusive $\gamma\gamma \to \tau\tau$ \cite{CMS:2024qjo} and the ATLAS analysis of the Drell-Yan tail \cite{ATLAS:2025oiy}—have pushed the experimental precision to the $\mathcal{O}(10^{-3})$ level. While $pp$ data offer superior statistics, they require a careful evaluation of EFT validity when probing the TeV scale. 
From a theoretical perspective, the SMEFT approach highlights that $a_\tau$ is not an isolated observable, but part of a broader structure dictated by electroweak gauge symmetry. Correlations with $Z$ and $W$ interactions imply that future constraints will benefit from global analyses combining multiple processes.

Looking ahead, significant progress is expected from Belle II and future facilities such as the FCC. These experiments will extend both precision and energy reach, bringing us closer to the Standard Model prediction of $a_\tau^{\text{SM}} = 117\,721 \, (5) \times 10^{-8}$. Belle II and the proposed FCC-ee are the most immediate candidates to reach sensitivities in the $10^{-5}$ range \cite{Bodrov:2024wrw, Buttazzo:2026amk, Dam:2018rfz}. In the longer term, the high-energy Muon Collider represents a further ambitious frontier that could push the resolution towards $\mathcal{O}(10^{-6})$ \cite{Wang:2024bfc, Buttazzo:2026amk}, provided its preliminary design challenges are overcome. In contrast, the FCC-hh stage in PbPb mode, while providing a unique UPC environment, is projected to yield more moderate exclusion limits at the $\mathcal{O}(10^{-2})$ level \cite{Inan:2026byu}.

In conclusion, the study of $a_{\tau}$ is transitioning from a largely unconstrained observable to a quantitatively accessible one. The synergy between the clean environment of UPCs and the high-energy reach of $pp$ and future colliders will be essential to determine whether the tau lepton reveals signatures of New Physics or confirms the Standard Model with unprecedented precision.

The measurement of $a_\tau$ thus represents a unique bridge between the precision and energy frontiers, with UPCs and high-energy pp collisions providing complementary and necessary information on the structure of the tau electromagnetic vertex.

\section*{Acknowledgement}

This work originated from and greatly benefited from the discussions held during the CERN Collider Cross Talk ``tau g-2 in HIN''. The author would like to express sincere gratitude to the organizers for the invitation and for providing such a stimulating environment for scientific exchange. Special thanks are also due to V. Cavasinni, C. Roda, and M. Verducci for their collaboration in research work relevant to this review.
This work is supported by ICSC – Centro Nazionale di Ricerca in High Performance Computing, Big Data and Quantum Computing, funded by European Union – NextGenerationEU, reference code CN\_00000013.


\begin{thebibliography}{99}
\bibitem{Eidelman:2007sb}
S.~Eidelman and M.~Passera,
Mod. Phys. Lett. A \textbf{22}, 159-179 (2007)
doi:10.1142/S0217732307022694
[arXiv:hep-ph/0701260 [hep-ph]].

\bibitem{Moroi:1995yh}
T.~Moroi,
Phys. Rev. D \textbf{53}, 6565-6575 (1996)
[erratum: Phys. Rev. D \textbf{56}, 4424 (1997)]
doi:10.1103/PhysRevD.53.6565
[arXiv:hep-ph/9512396 [hep-ph]].

\bibitem{Crivellin:2021spu}
A.~Crivellin, M.~Hoferichter and J.~M.~Roney,
Phys. Rev. D \textbf{106}, no.9, 093007 (2022)
doi:10.1103/PhysRevD.106.093007
[arXiv:2111.10378 [hep-ph]].

\bibitem{DAlise:2024qmp}
A.~D'Alise, G.~Fabiano, D.~Frattulillo, D.~Iacobacci, F.~Sannino, P.~Santorelli and N.~Vignaroli,
Nucl. Phys. B \textbf{1006}, 116631 (2024)
doi:10.1016/j.nuclphysb.2024.116631
[arXiv:2403.17614 [hep-ph]].

\bibitem{Vignaroli:2019lkg}
N.~Vignaroli,
Nuovo Cim. C \textbf{43}, no.2-3, 53 (2020)
doi:10.1393/ncc/i2020-20053-0
[arXiv:1912.00899 [hep-ph]].

\bibitem{Vignaroli:2018lpq}
N.~Vignaroli,
Phys. Rev. D \textbf{99}, no.3, 035021 (2019)
doi:10.1103/PhysRevD.99.035021
[arXiv:1808.10309 [hep-ph]].

\bibitem{Grzadkowski:2010es}
B.~Grzadkowski, M.~Iskrzynski, M.~Misiak and J.~Rosiek,
JHEP \textbf{10}, 085 (2010)
doi:10.1007/JHEP10(2010)085
[arXiv:1008.4884 [hep-ph]].

\bibitem{Brivio:2017vri}
I.~Brivio and M.~Trott,
Phys. Rept. \textbf{793}, 1-98 (2019)
doi:10.1016/j.physrep.2018.11.002
[arXiv:1706.08945 [hep-ph]].

\bibitem{FCC:2018evy}
A.~Abada \textit{et al.} [FCC],
Eur. Phys. J. ST \textbf{228}, no.2, 261-623 (2019)
doi:10.1140/epjst/e2019-900045-4

\bibitem{Vignaroli:2025pwn}
N.~Vignaroli,
JHEP \textbf{10}, 125 (2025)
doi:10.1007/JHEP10(2025)125
[arXiv:2507.01130 [hep-ph]].

\bibitem{Frigerio:2024jlh}
M.~Frigerio and N.~Vignaroli,
JHEP \textbf{04}, 008 (2025)
doi:10.1007/JHEP04(2025)008
[arXiv:2409.02721 [hep-ph]].

\bibitem{Abdallah:2003xd}
J.~Abdallah \textit{et al.} [DELPHI],
Eur. Phys. J. C \textbf{35}, 159-170 (2004)
doi:10.1140/epjc/s2004-01852-y
[arXiv:hep-ex/0406010 [hep-ex]].

\bibitem{CMS:2024qjo}
A.~Hayrapetyan \textit{et al.} [CMS],
Rept. Prog. Phys. \textbf{87}, no.10, 107801 (2024)
doi:10.1088/1361-6633/ad6fcb
[arXiv:2406.03975 [hep-ex]].

\bibitem{ATLAS:2025oiy}
G.~Aad \textit{et al.} [ATLAS],
JHEP \textbf{10}, 054 (2025)
doi:10.1007/JHEP10(2025)054
[arXiv:2503.19836 [hep-ex]].

\bibitem{Fermi:1924tc}
E.~Fermi,
Z. Phys. \textbf{29}, 315-327 (1924)
doi:10.1007/BF03184853

\bibitem{vonWeizsacker:1934nji}
C.~F.~von Weizsacker,
Z. Phys. \textbf{88}, 612-625 (1934)
doi:10.1007/BF01333110

\bibitem{Williams:1934ad}
E.~J.~Williams,
Phys. Rev. \textbf{45}, 729-730 (1934)
doi:10.1103/PhysRev.45.729

\bibitem{Budnev:1975poe}
V.~M.~Budnev, I.~F.~Ginzburg, G.~V.~Meledin and V.~G.~Serbo,
Phys. Rept. \textbf{15}, 181-281 (1975)
doi:10.1016/0370-1573(75)90009-5

\bibitem{Vidovic:1992ik}
M.~Vidovic, M.~Greiner, C.~Best and G.~Soff,
Phys. Rev. C \textbf{47}, 2308-2319 (1993)
doi:10.1103/PhysRevC.47.2308

\bibitem{Beresford:2019gww}
L.~Beresford and J.~Liu,
Phys. Rev. D \textbf{102}, no.11, 113008 (2020)
[erratum: Phys. Rev. D \textbf{106}, no.3, 039902 (2022)]
doi:10.1103/PhysRevD.102.113008
[arXiv:1908.05180 [hep-ph]].

\bibitem{Verducci:2023cgx}
M.~Verducci, C.~Roda, V.~Cavasinni and N.~Vignaroli,
Phys. Rev. D \textbf{110}, no.5, 052001 (2024)
doi:10.1103/PhysRevD.110.052001
[arXiv:2307.15160 [hep-ph]].

\bibitem{Dyndal:2020yen}
M.~Dyndal, M.~Klusek-Gawenda, M.~Schott and A.~Szczurek,
Phys. Lett. B \textbf{809}, 135682 (2020)
doi:10.1016/j.physletb.2020.135682
[arXiv:2002.05503 [hep-ph]].

\bibitem{ATLAS:2022ryk}
G.~Aad \textit{et al.} [ATLAS],
Phys. Rev. Lett. \textbf{131}, no.15, 151802 (2023)
doi:10.1103/PhysRevLett.131.151802
[arXiv:2204.13478 [hep-ex]].

\bibitem{CMS:2022arf}
A.~Tumasyan \textit{et al.} [CMS],
Phys. Rev. Lett. \textbf{131}, 151803 (2023)
doi:10.1103/PhysRevLett.131.151803
[arXiv:2206.05192 [nucl-ex]].

\bibitem{Dittmaier:2025ikh}
S.~Dittmaier, T.~Engel, J.~L.~H.~Ariza and M.~Pellen,
JHEP \textbf{08}, 051 (2025)
doi:10.1007/JHEP08(2025)051
[arXiv:2504.11391 [hep-ph]].

\bibitem{LoChiatto:2024dmx}
P.~Lo Chiatto,
Phys. Rev. D \textbf{112}, no.1, 015017 (2025)
doi:10.1103/8gtq-twfc
[arXiv:2408.04553 [hep-ph]].

\bibitem{Shao:2023bga}
D.~Shao, B.~Yan, S.~R.~Yuan and C.~Zhang,
Sci. China Phys. Mech. Astron. \textbf{67}, no.8, 281062 (2024)
doi:10.1007/s11433-024-2389-y
[arXiv:2310.14153 [hep-ph]].

\bibitem{Beresford:2024dsc}
L.~Beresford, S.~Clawson and J.~Liu,
Phys. Rev. D \textbf{110}, no.9, 092016 (2024)
doi:10.1103/PhysRevD.110.092016
[arXiv:2403.06336 [hep-ph]].

\bibitem{Bruce:2018yzs}
R.~Bruce, D.~d'Enterria, A.~de Roeck, M.~Drewes, G.~R.~Farrar, A.~Giammanco, O.~Gould, J.~Hajer, L.~Harland-Lang and J.~Heisig, \textit{et al.}
J. Phys. G \textbf{47}, no.6, 060501 (2020)
doi:10.1088/1361-6471/ab7ff7
[arXiv:1812.07688 [hep-ph]].

\bibitem{Belle-II:2018jsg}
E.~Kou \textit{et al.} [Belle-II],
PTEP \textbf{2019}, no.12, 123C01 (2019)
[erratum: PTEP \textbf{2020}, no.2, 029201 (2020)]
doi:10.1093/ptep/ptz106
[arXiv:1808.10567 [hep-ex]].


\bibitem{Bodrov:2024wrw}
D.~Bodrov,
Int. J. Mod. Phys. A \textbf{39}, no.26n27, 2442006 (2024)
doi:10.1142/S0217751X24420065
[arXiv:2405.16512 [hep-ex]].


\bibitem{Bernabeu:2008ii}
J.~Bernabeu, G.~A.~Gonzalez-Sprinberg and J.~Vidal,
JHEP \textbf{01}, 062 (2009)
doi:10.1088/1126-6708/2009/01/062
[arXiv:0807.2366 [hep-ph]].

\bibitem{Gogniat:2026zvf}
J.~Gogniat, M.~Hoferichter and G.~Levati,
[arXiv:2604.16598 [hep-ph]].

\bibitem{Buttazzo:2026amk}
D.~Buttazzo, G.~Levati, Y.~Ma, F.~Maltoni, P.~Paradisi and Z.~Wang,
[arXiv:2604.14281 [hep-ph]].


\bibitem{Dam:2018rfz}
M.~Dam,
SciPost Phys. Proc. \textbf{1}, 041 (2019)
doi:10.21468/SciPostPhysProc.1.041
[arXiv:1811.09408 [hep-ex]].

\bibitem{Inan:2026byu}
S.~C.~{\.I}nan and A.~V.~Kisselev,
[arXiv:2601.18288 [hep-ph]].

\bibitem{InternationalMuonCollider:2025sys}
C.~Accettura \textit{et al.} [International Muon Collider],
[arXiv:2504.21417 [physics.acc-ph]].


\bibitem{Denizli:2024uwv}
H.~Denizli, A.~Senol and M.~K{\"o}ksal,
Chin. J. Phys. \textbf{95}, 1250-1258 (2025)
doi:10.1016/j.cjph.2025.04.020
[arXiv:2408.16106 [hep-ph]].


\bibitem{Wang:2024bfc}
Z.~Wang,
PoS \textbf{ICHEP2024}, 327 (2025)
doi:10.22323/1.476.0327
[arXiv:2410.12663 [hep-ph]].

 \bibitem{Vignaroli:in-preparation}
 N. Vignaroli, in preparation.

\end{thebibliography}
\end{document}